# Tomonaga-Luttinger liquid and charge-density wave in a quasi-one-dimensional material

**Authors:** Jing Li[1]*, Guo-Wei Yang[1,2]*, Bai-Zhuo Li[3], Yi Liu[4], Si-Qi Wu[5], Ji-Yong Liu[6], Jin-Ke Bao[7], Xiaoxian Yan[8], Hua-Xun Li[1], Jia-Xin Li[1], Jia-Lu Wang[7], Yun-Lei Sun[9], Yi-Ming Lu[1], Jia-Yi Lu[1], Yi-Qiang Lin[1], Hui Xing[8], Chao Cao[1,2]†, Hao Jiang[10]†, Yang Liu[1,2]†, Guang-Han Cao[1,11]†, Hai-Qing Lin[1,12]

**Affiliations:**
[1]School of Physics, Zhejiang University, Hangzhou 310058, China.
[2]Center for Correlated Matter and School of Physics, Zhejiang University, Hangzhou 310058, China.
[3]School of Physics and Optoelectronic Engineering, Shandong University of Technology, Zibo 255000, China.
[4]Department of Applied Physics, Zhejiang University of Technology, Hangzhou 310023, China.
[5]Department of Physics, Hong Kong University of Science and Technology, Clear Water Bay, Hong Kong, China.
[6]Department of Chemistry, Zhejiang University, Hangzhou 310058, China.
[7]School of Physics, Hangzhou Normal University, Hangzhou 310036, China.
[8]Key Laboratory of Artificial Structures and Quantum Control, and Shanghai Center for Complex Physics, School of Physics and Astronomy, Shanghai Jiao Tong University, Shanghai 200240, China.
[9]School of Information and Electrical Engineering, Hangzhou City University, Hangzhou 310015, China.
[10]School of Physics and Optoelectronics, Xiangtan University, Xiangtan 411105, China.
[11]Institute of Fundamental and Transdisciplinary Research, and State Key Laboratory of Silicon and Advanced Semiconductor Materials, Zhejiang University, Hangzhou 310058, China.
[12]Institue of Advanced Study in Physics, Zhejiang University, Hangzhou 310058, China.
†Corresponding authors. E-mail: ccao@zju.edu.cn; hjiang@xtu.edu.cn; yangliuphys@zju.edu.cn; ghcao@zju.edu.cn;
*These authors contributed equally.

**In one-dimensional (1D) electron systems, the Fermi liquid state breaks down due either to electron interactions, which results in a Tomonaga-Luttinger liquid (TLL) state, or to Peierls instability, which leads to an insulating charge-density-wave (CDW) phase. In general, these two phenomena are mutually exclusive, and their coexistence remains elusive in real materials. Here, we report the discovery of a new quasi-1D material, $Cs_{1-\delta}Cr_3S_3$, which unexpectedly exhibits coexistence of the antithetical CDW and TLL states. The CDW state is evidenced by the intra-unit-cell dimerization and the opening of an optical band gap of ~250 meV. Meanwhile, TLL behaviour is unambiguously demonstrated by the measurements of electrical transport and angle-resolved photoemission spectroscopy, which reveals power-law scaling with temperature, bias voltage, and electron energy. Band structure calculations reveal isolated, linearly dispersive, 1D bands around the Fermi level. For the dimerized CDW phase, the 1D Fermi-surface sheets located at the boundary of the Brillouin zone are gapped from intra-unit-cell bond symmetry breaking. Experimentally, subtle Cs vacancies shift the Fermi level into the linearly dispersive valence band, enabling the observation of TLL behaviour without interrupting the CDW order. This work establishes $Cs_{1-\delta}Cr_3S_3$ as a rare material platform in which the antagonistic Fermi-liquid instabilities coexist and intertwine, opening new avenues for studying emergent quantum phenomena in 1D systems.**

TLL states and Peierls distortion, arising from electron interactions and electron–phonon coupling respectively, represent an apparently irreconcilable electronic dichotomy in 1D material systems. TLL emerges from collective excitations in a 1D chain, i.e., bosonic excitations of spinons and holons, associated with linear band dispersion at the Fermi level[1-5]. Yet, the same 1D confinement also favours a Peierls distortion, which breaks translational symmetry and opens a band gap, driving the system into an insulating CDW state[6-8]. This intrinsic antagonism has long rendered their coexistence elusive in real materials.

Quasi-1D materials normally host either TLL or CDW order, as exemplified by Bechgaard salts[9], purple bronzes[10,11], spin chain compounds[12], Cr-based arsenides[13-16], quantum wires[17,18], carbon nanotubes[19,20], and moiré $WTe_2$[21]. A broader survey (Extended Data Table 1) reveals the rare coexistence of the two states. Here we report a new quasi-1D compound, $Cs_{1-\delta}Cr_3S_3$, in which TLL behaviour emerges in the CDW phase.

**Peierls structural distortion**

Single crystals of $Cs_{1-\delta}Cr_3S_3$ were grown by a flux growth method (see Methods). The as-grown crystals are needle-like, silver-lustrous, and stable in air. The chemical composition was determined by energy-dispersive X-ray spectroscopy (EDS), which shows systematic Cs deficiency with $\delta = 0.044 \pm 0.015$ for samples studied in the present work (Extended Data Figs. 1a, b). Single-crystal X-ray diffraction (SXRD) identifies an alternative crystal structure analogous to $AkCr_3As_3$[16] and $Ak_2Mo_6Ch_6$[22] with space group $P6_3/m$ (No. 176) (Extended Data Table 2). The key structural unit is the 1D $Cr_3S_3$ double-walled subnanotube, the inside of which forms face-sharing $Cr_6$ octahedra (Fig. 1a and Extended Data Figs. 1c, d). These coaxial tubes extend infinitely along the crystallographic $c$ direction and are spatially separated by $Cs^+$ cations, making strong one dimensionality.

As is well known, quasi-1D systems are highly susceptible to Peierls instability, which tends to induce lattice dimerization[23]. In the present case of $CsCr_3S_3$, the lattice dimerization (Fig. 1b) with the individual atomic displacements shown in Extended Data Fig. 1e leads to a lower space group $P\bar{3}$ (No. 147). Our density functional theory (DFT) calculations of phonon spectra reveal that the undimerized $CsCr_3S_3$ has enormous imaginary modes in the $k_z = 0$ plane, indicating strong dynamical instability (Fig. 1c). In contrast, the dimerized structure relieves such an instability (Fig. 1d) as well as lowers the total energy by ~2 meV/f.u. (Extended Data Fig. 1f), suggesting that the dimerized CDW phase may be realized. The fingerprint of the dimerized phase is the appearance of (001) reflection, which is extinct in the undimerized structure due to the screw symmetry ($6_3$) along chains (Figs. 1e, f). Indeed, the tiny extra (001) reflection was easily observed in the SXRD patterns with elongated crystals (Fig. 1g), although it seemed too weak to be detected for short crystals (Extended Data Fig. 2a). The whole set of the SXRD data was also refined using the dimerized structure. However, the resulting lattice distortion (Extended Data Table 3) is much less prominent than that obtained through DFT optimization. This is probably due to the presence of multiple domains, as well as the negligible contribution of the forbidden reflections to the SXRD refinement.

In order to confirm the structural dimerization for the $Cs_{1-\delta}Cr_3S_3$ crystals, we made micro-area analysis utilizing high-resolution transmission electron microscopy. As shown in Fig. 1h, the selected-area electron diffraction (SAED) at room temperature reveals significant (00$l$) reflections, with $l$ = odd numbers, expected for the dimerized structure (see simulated SAED patterns in Extended Data Figs.

2d, e). Notably, the uniformity of SAED patterns across multiple spatial regions demonstrates that the modulation is homogeneous at nanoscale without signature of phase separation (Extended Data Fig. 2b). Such (00$l$) reflections can be clearly observed even for the highly Cs-deficient ($\delta \approx 0.38$) sample (Extended Data Fig. 2c). This result indicates the robustness of Peierls structural distortion against the Cs deficiency. The Peierls transition temperature in $Cs_{1-\delta}Cr_3S_3$ is expected to be much higher than room temperature, because no anomaly is observed in the temperature dependence of magnetic susceptibility up to 400 K (Extended Data Fig. 3).

The Peierls transition normally induces a direct band gap which can be detected by an optical absorption measurement. As illustrated in Fig. 1i, the optical spectrum of the $Cs_{1-\delta}Cr_3S_3$ crystal exhibits a finite absorption edge, indicating a direct optical gap of ~250 meV. Measurements on additional samples are presented in Extended Data Fig. 2f, which corroborate the gap opening associated with the structural distortion.

**Transport and thermodynamic signatures of TLL**
We next turn to the transport properties of $Cs_{1-\delta}Cr_3S_3$ single crystals. The temperature-dependent resistivity, $\rho(T)$, shows a semiconducting-like behaviour (Fig. 2a). Above ~20 K, the $\rho(T)$ data severely deviates from the Arrhenius law expected for a conventional gapped semiconductor, $\rho(T) \propto \exp[E_a/(k_BT)]$, where $E_a$ denotes the activation energy. Instead, these data follow a power-law dependence, $\rho(T) \propto T^{-\alpha}$ with $\alpha = 1.9$, a characteristic of TLL transport in interacting 1D systems[24,25]. On the other hand, the low-temperature (below ~15 K) data follow the Arrhenius relation, with an $E_a$ value of 1.77 meV, much smaller than the Peierls gap inferred from the above optical measurement. This charge gap opening in the TLL regime can be explained in terms of an umklapp scattering near a commensurate filling[26], since the valence band is nearly fully filled in $Cs_{1-\delta}Cr_3S_3$ (see below). Note that a similar crossover in electrical transport is observed in carbon nanotube, although the underlying mechanism is different[27-29].

For the putative TLL state, current–voltage (I–V) measurements, which probe the theoretically predicted crossover from ohmic to power-law behaviour, provides a stringent and model-independent check[30]. Fig. 2b displays a series of isothermal I–V curves of $Cs_{1-\delta}Cr_3S_3$ from 15 K to 260 K. At high temperatures and low bias voltages, the curves exhibit conventional ohmic behaviour. Beyond a critical bias that scales with temperature, they systematically deviate from Ohm's law, and evolve into a power-law form, $I \propto V^{\varphi+1}$, with a $\varphi$-value of ~1.54 (dashed line). To quantitatively capture this evolution, we extracted the power exponent $\varphi$ over a wide range of temperatures and voltages (Fig. 2c). The result exhibits a clear crossover from ohmic to non-ohmic in the vicinity of $\eta eV \approx 2.4 k_B T$, where $\eta$ represents the ratio of the junction voltage to the applied one, $k_B$ the Boltzmann constant, and $e$ the electron charge. Within the non-ohmic regime, the average value of $\varphi$ is 1.54.

In the TLL framework, the I–V response can be universally expressed by[31],

$$I = I_0 T^{\alpha+1} \sinh\left(\frac{\eta eV}{2k_BT}\right) \left|\Gamma\left(1 + \frac{\varphi}{2} + i\frac{\eta eV}{2k_BT}\right)\right|^2.$$

Here, $I_0$ and $\eta$ are constants independent of temperature and voltage, and $\Gamma(x)$ corresponds to gamma function. The formula indicates that the electrical transport is governed by a universal function of the reduced variable $eV/k_BT$, with asymptotic limits of $\rho \propto T^{-\alpha}$ in the thermal regime and $I \propto V^{\varphi+1}$ in the field-driven regime. Notably, the exponents $\alpha$ and $\varphi$, extracted from independent measurements, are basically consistent since $\varphi = \alpha$ in a clean TLL[17]. When rescaled accordingly, I–V curves collapse onto a single master curve (Fig. 2d), corroborating the TLL transport behaviour.

Measurements of heat capacity and thermal transport further support the existence of a TLL state in $Cs_{1-\delta}Cr_3S_3$. Fitting of the low-temperature specific-heat data by formula $C/T = \gamma + \beta T^2$ yields a sizeable $\gamma$ value of 3.43 mJ mol$^{-1}$ K$^{-2}$ (Fig. 2e). The result is incompatible with a gapped CDW insulator in which the linear term $\gamma T$ is expected to be zero. Instead, it points to a spin-charge-separated TLL state, in which spinons and holons effectively contribute a linear specific heat of $(v_F/\mu_\rho + v_F/\mu_\sigma)\gamma_0 T/2$ (ref. 25), where $v_F$ is the Fermi velocity, $\mu_\rho$ the velocity of holons (charge excitations), $\mu_\sigma$ the velocity of spinons (spin excitations), and $\gamma_0 = \pi^2 k_B^2 N(E_F)/3$ denotes the expected specific-heat coefficient derived from the bare electronic density of states at the Fermi level, $N(E_F)$. Using the calculated $N(E_F)$ value of 0.69 eV$^{-1}$ f.u.$^{-1}$ for the dimerized $Cs_{1-\delta}Cr_3S_3$ (see Methods), the $v_F/\mu_\sigma$ value is estimated to be 4.2, with the charge-sector contribution suppressed by the charge gap mentioned above. This implies that the velocity of spinons is of the same order of magnitude as the Fermi velocity, $v_F = 4.65 \times 10^5$ m/s, obtained from band-structure calculations. The low-temperature thermal conductivity data (Fig. 2f) signifies a substantial non-phononic contribution by $\kappa/T = A + BT^2$. Thus the Lorentz number $\kappa/(\sigma T)$, where $\sigma$ is the electrical conductivity, is orders of magnitude larger than that of a Fermi liquid, indicating the breakdown of Wiedemann-Franz law. The linear term coefficient ($A = 2.4$ W m$^{-1}$ K$^{-2}$) is basically consistent with the theoretically expected cumulative value from the quantized thermal conductance of a backscattering-free LL state[32], $A \approx 5.3$ W m$^{-1}$ K$^{-2}$ (detailed in Methods).

**ARPES signatures of TLL**

To investigate the quasi-1D nature of electronic states in $Cs_{1-\delta}Cr_3S_3$, we performed systematic 3D mapping in the reciprocal space using synchrotron-based angle-resolved photoemission spectroscopy (ARPES). $Cs_{1-\delta}Cr_3S_3$ single crystals were cleaved *in situ* with the c axis aligned parallel to the detector slit, corresponding to the A–Γ–A path (denoted as $k_z$ in Fig. 3a). By varying the photon energy ($h\nu$) and performing deflective angular scans, we obtained the momentum dispersion over the entire Brillouin zone. We found that the constant energy contours near $E_F$ consist of flat sheets, which cut through the A–L–H plane and exhibit no noticeable $k_x$ or $k_y$ dispersion, confirming its 1D nature. As such, we focus on a representative $E$–$k_z$ spectrum to capture its essential electronic properties.

Figure 3b presents the A–Γ–A spectrum, and its second derivative (Fig. 3c) highlights the band dispersions. Importantly, only a single set of bands cross $E_F$, which therefore dominates the low-energy physics in $Cs_{1-\delta}Cr_3S_3$. The high-resolution spectrum (Fig. 3d) demonstrates its line dispersion with a continuous suppression of spectral weight near $E_F$. The linear dispersion is a key prerequisite of TLL theory, and the observed spectral-weight suppression serves as its characteristic manifestation[14]. These observations reveal fingerprints of the TLL electronic state in $Cs_{1-\delta}Cr_3S_3$.

We next examine these two TLL signatures quantitatively. The band dispersions were extracted from momentum distribution curves (MDCs) at different energies (Fig. 3e). Despite the reduced spectral intensity near $E_F$, all four branches of bands exhibit robust linear dispersions over a wide energy range. Their extrapolated intersections lie slightly above $E_F$ (~40 meV), consistent with a minor hole doping by the ~5% Cs deficiency. To verify this subtle phenomenon, we further conducted *in-situ* electron doping via potassium deposition. The valence bands indeed move downward after electron doping, clearly revealing a dimerization-induced gap (Extended Data Fig. 4).

To verify the characteristic power-law suppression of spectral weight in TLL, we performed temperature-dependent measurements (Extended Data Fig. 5) and extracted energy distribution curves (EDCs) at 20–120 K, as shown in Fig. 3f. Unlike the sharp Fermi edge near $E_F$ in higher-dimensional Fermi liquids (e.g., polycrystalline Ag), the density of states (DOS) for $Cs_{1-\delta}Cr_3S_3$ smoothly vanishes near $E_F$. At finite temperatures, the TLL spectral intensity follows[33],

$$I(\varepsilon, T) \propto T^\alpha \cosh(\frac{\varepsilon}{2}) \left|\Gamma\left(\frac{1+\alpha}{2} + i\frac{\varepsilon}{2\pi}\right)\right|^2 f(\varepsilon),$$

where $\varepsilon = E/k_B T$ is the scaled energy and $f(\varepsilon)$ is the Fermi-Dirac function. This relation implies that $I(\varepsilon)/T^\alpha$ follows a universal scaling law with respect to $\varepsilon$, so that the rescaled EDCs should collapse onto a single curve at the correct $\alpha$. By testing different values (Extended Data Fig. 6), we find that the optimal $\alpha$ value is about 1.2. Taken together, the observations of the linear band dispersion near $E_F$ and the temperature-independent scaling behaviour provide unambiguous evidence for the TLL ground state in $Cs_{1-\delta}Cr_3S_3$.

**Understanding of TLL state in a CDW phase**

To understand the experimental results above, which show a seemingly conflicting phenomenon— existence of TLL state in a CDW phase, we first performed DFT calculations on the undimerized and dimerized $CsCr_3S_3$. The results show that both phases exhibit pronounced dispersions along $k_z$ (with a band width $W_z$), but the dispersions within the $k_x-k_y$ plane (with a band width $W_{xy}$) is negligible at around $E_F$ (Figs. 4a, b). This corresponds to a very large anisotropy parameter, $\lambda = W_z/W_{xy} \approx 173$, indicating exceedingly weak interchain hopping. The extremely 1D electronic bands with linear dispersion near $E_F$ are verified by the ARPES measurements above. In the undimerized phase, the system is gapless, with the Fermi surface touching the Brillouin zone (BZ) boundary, i.e., the A–L–H–A plane. According to traditional Peierls scenario, the perfect Fermi-surface nesting at $q = 2k_F$ leads to the opening of a direct gap of 197 meV (Figs. 4c, d). In the present case, however, the Peierls instability roots in intra-unit-cell bond symmetry breaking, i.e., bond disproportionation within the existing unit cell without any change in translational periodicity (Extended Data Fig. 9, see Methods for the details). To model the Cs deficiency observed experimentally, we applied virtual-crystal approximation (VCA), which indicates that the Cs deficiency induces a hole doping. This doping shifts the Fermi level into the valence band, which leads to an increased optical gap of ~250 meV probed by the optical measurement above (right panel of Fig. 4i).

With the aim of uncovering the microscopic origin of linear bands, we constructed a minimal tight-binding (TB) model for the 1D $Cr_6$ subnanotubes (see Methods for the details). The model comprises of two bands using two molecular orbitals (MOs) as the basis set, each of which is a linear combination of three Cr-$d_{yz/xz}$ orbitals (Figs. 4e, f). The TB model successfully reproduces the DFT bands, in particular, giving rise to the linear dispersions near $E_F$, with a slope analytically given by $t_1-3t_3$ as the leading order (Figs. 4g, h and Extended Data Fig. 7). Higher-order hopping terms are also included in the Hamiltonian: the odd-neighbour hoppings $t_{2n+1}$ enter the off-diagonal matrix elements and contribute additional linear dispersions with a correction factor $-(n+1)^2+n^2$, whereas the even-neighbour hoppings $t_{2n}$ enter the diagonal elements and lead to quadratic corrections. Our result shows that the nearest-neighbour hopping makes a dominant contribution to the linear dispersion (Extended Data Tables 4, 5). Such a MO-based symmetry-protected linearity provides the microscopic basis for the TLL state.

Taken together, our results outline a coherent mechanism that reconciles the seemingly conflicting observations (Fig. 4i). In $Cs_{1-\delta}Cr_3S_3$, the $Cr_6$ subnanotubes undergo lattice dimerization accompanied with an opening of the Peierls gap, irrespective of the hole doping. Upon hole doping, $E_F$ moves into the lower MO-based bonding band where the linear dispersion remains, giving rise to the TLL behaviours in the symmetry-broken CDW phase. Notably, the hole doping is away from the 1D $Cr_6$ subnanotubes, which induces insignificant disorder into the $Cr_6$ subnanotubes due to the structural protection by the $S_6$ nanotubes.

## Concluding remarks

To summarize, we have discovered a new quasi-1D material $Cs_{1-\delta}Cr_3S_3$ which unexpectedly exhibits both CDW order and TLL behaviour below room temperature. The CDW order, being robust against hole doping due to the Cs deficiency, is characterized by the commensurate intra-unit-cell dimerization, which opens a band gap of ~200 meV. The TLL state is supported by the specific-heat, thermal conductivity and ARPES measurements, and is manifested by power-law scaling with temperature, bias voltage, and electron energy, which gives an exponent of $\alpha = 1.2–1.9$. Using the relation $\alpha = \frac{K+K^{-1}-2}{4}$ (ref. 4 and ref. 24), the Tomonaga–Luttinger parameter $K$ can be obtained, offering a universal metric for electron interactions in a TLL state. Generally speaking, smaller $K$ value ($K < 1$) means stronger repulsive interactions. In $Cs_{1-\delta}Cr_3S_3$, remarkably, different probes yield a similar $K$ value of 0.11–0.15, which is among the lowest in known TLL states (Extended Data Table 1). This underscores the strong repulsive interactions in the present system, likely associated with the inherent strong electron correlations in Cr-based systems[14,34]. The strongly repulsive TLL persists within the CDW phase, creating a hybrid quantum state that blends paradoxical phenomena in ways not achievable in most 1D systems. The crucial point here is that the hole doping does not destroy the commensurate CDW order, but rather induces a TLL state in $Cs_{1-\delta}Cr_3S_3$. Our work provides a conceptual blueprint for realizing novel quantum states in 1D electron system.

# Figures

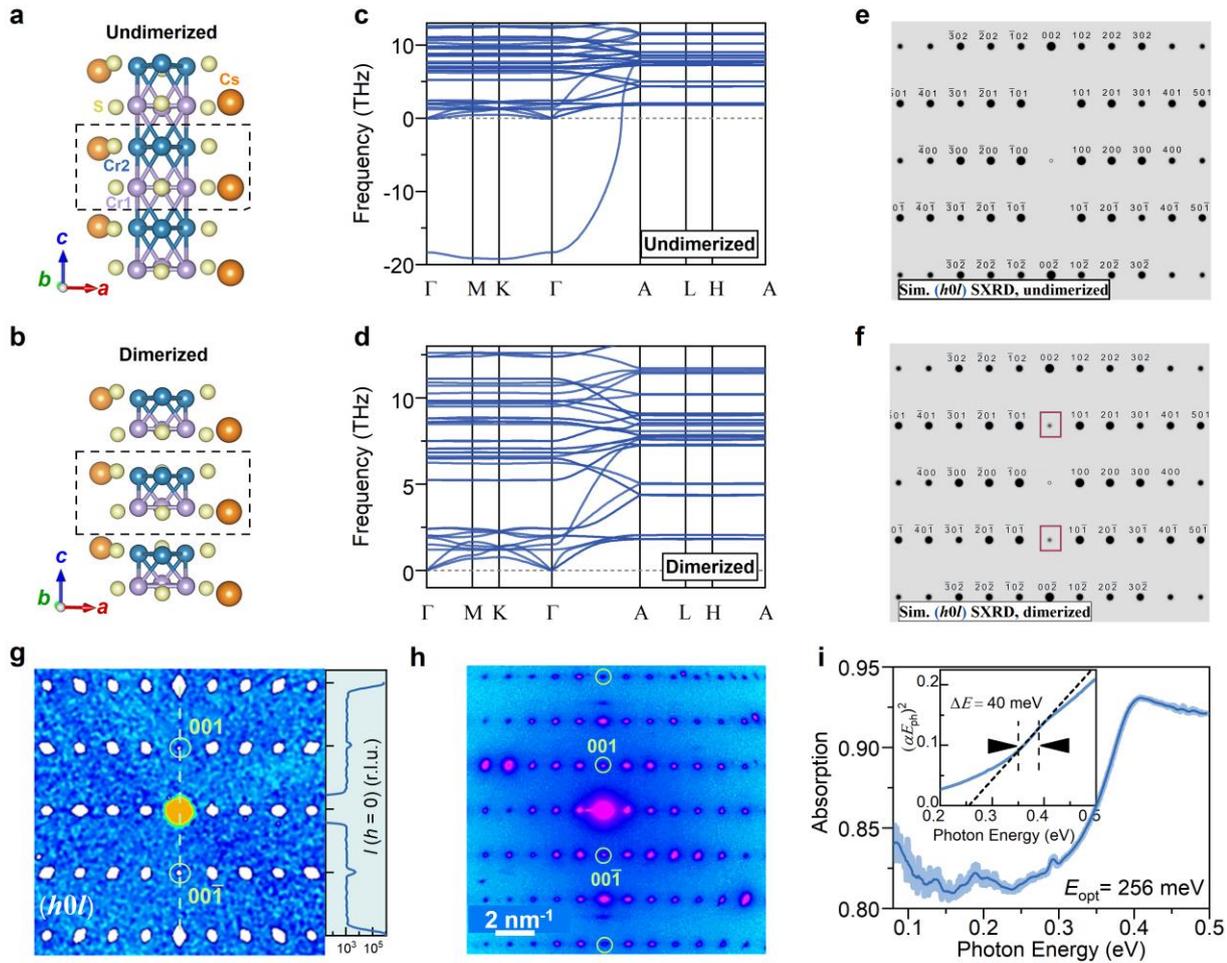

**Figure 1. Peierls structural phase transition in Cs$_{1-\delta}$Cr$_3$S$_3$. a, b,** Side views of the crystal structure before and after dimerization. In the dimerized phase, Cr displacements along *c* direction are exaggerated to emphasize the dimerization of the Cr chain. **c, d,** Phonon spectra for undistorted and distorted structures respectively. **e, f,** Simulated (Sim.) SXRD diffraction pattern of the (*h*0*l*) reciprocal plane of two structures (DFT relaxed) with Mo $K_\alpha$ radiation. **g,** Experimental SXRD diffraction patterns (*h*0*l*) at 170 K. The right-side panel shows the line-cut data from (00$\bar{2}$) to (002) in $c^*$ direction, plotted in logarithmic scale. **h,** A selected area electron diffraction pattern in (*h*0*l*) plane measured at 300 K. **i,** Optical absorption spectrum measured at room temperature. The inset shows derivation of an optical energy gap by Tauc's method.

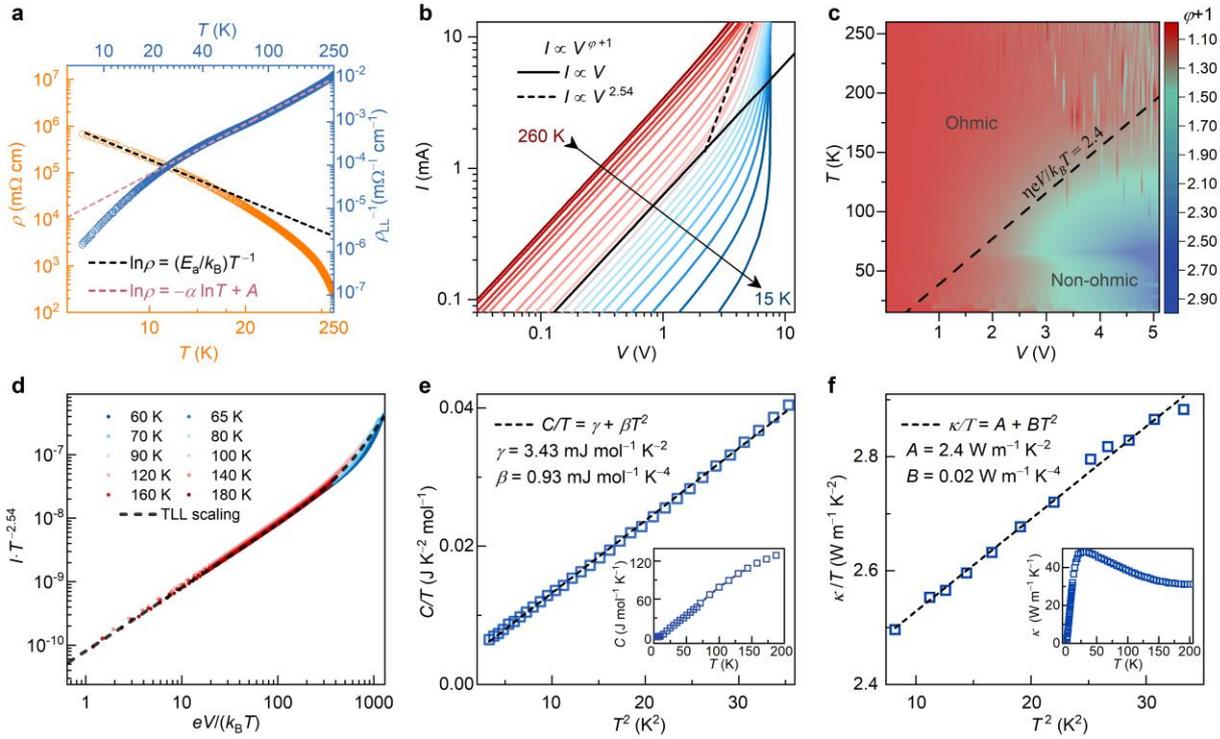

**Figure 2. Evidence of TLL from transport and thermal measurements for Cs$_{1-\delta}$Cr$_3$S$_3$. a,** Electrical resistivity as a function of temperature with current applied along the *c* axis. An a.c. excitation of 0.2 V was used to ensure ohmic response. **b,** *I–V* curves measured at different temperatures. The solid and dashed black lines represent a typical ohmic-to-nonohmic crossover. **c,** Power exponent ($\varphi+1$) plotted as a function of voltage and temperature, extracted from $I \propto V^{\varphi+1}$. **d,** Normalized *I–V* curves from 60 K to 180 K fitted by TLL theory with $\alpha = 1.54$ and $\eta = 0.008$. **e,** Temperature dependence of specific heat plotted as *C/T* versus $T^2$, used to extract the electronic ($\gamma$) and phononic ($\beta$) specific-heat coefficients. The inset shows the temperature dependence of specific heat. **f,** Thermal conductivity divided by temperature ($\kappa/T$) as a function of $T^2$ between 2.5 K and 6 K. The data are fitted by $\kappa/T = A + BT^2$, where *A* and *B* correspond to the electronic and phononic contributions, respectively. The inset shows the temperature dependence of thermal conductivity.

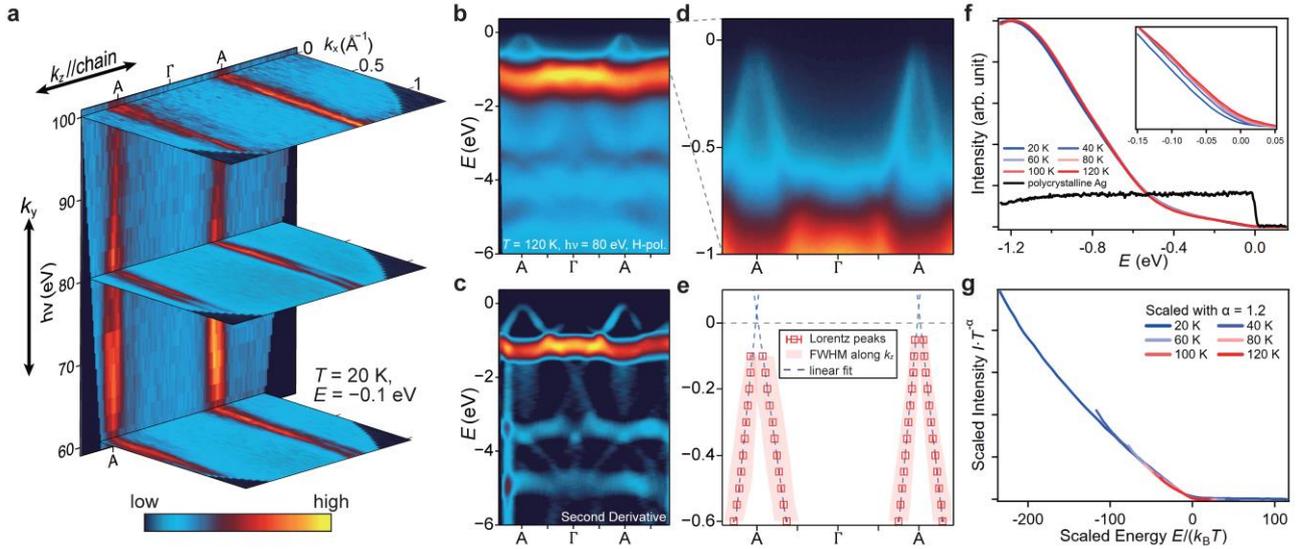

**Figure 3. ARPES evidence of TLL state in Cs$_{1-\delta}$Cr$_3$S$_3$. a,** Three-dimensional mapping of quasi-1D electronic bands. The map, corresponding to a constant energy of −0.1 eV, essentially consists of flat sheets locating near the A–L–H plane. The dispersion along $k_y$ is obtained by scanning the photon energy. **b,** The ARPES spectrum along A–Γ–A path taken with $h\nu$ = 80 eV. **c,** The second derivative plot of **b**. **d,** Detailed band dispersion near $E_F$. **e,** Band positions fitted from MDCs at different energies. The light red shaded regions represent the full width at half maximum (FWHM) of each band, and the blue dashed lines refer to the linear fit of band dispersion. **f,** EDCs integrated along A–Γ taken at different temperatures, compared with polycrystalline Ag as a reference. The inset is a close-up near $E_F$. **g,** Scaling of temperature-dependent EDCs according to the power law of TLL, with $\alpha$ = 1.2.

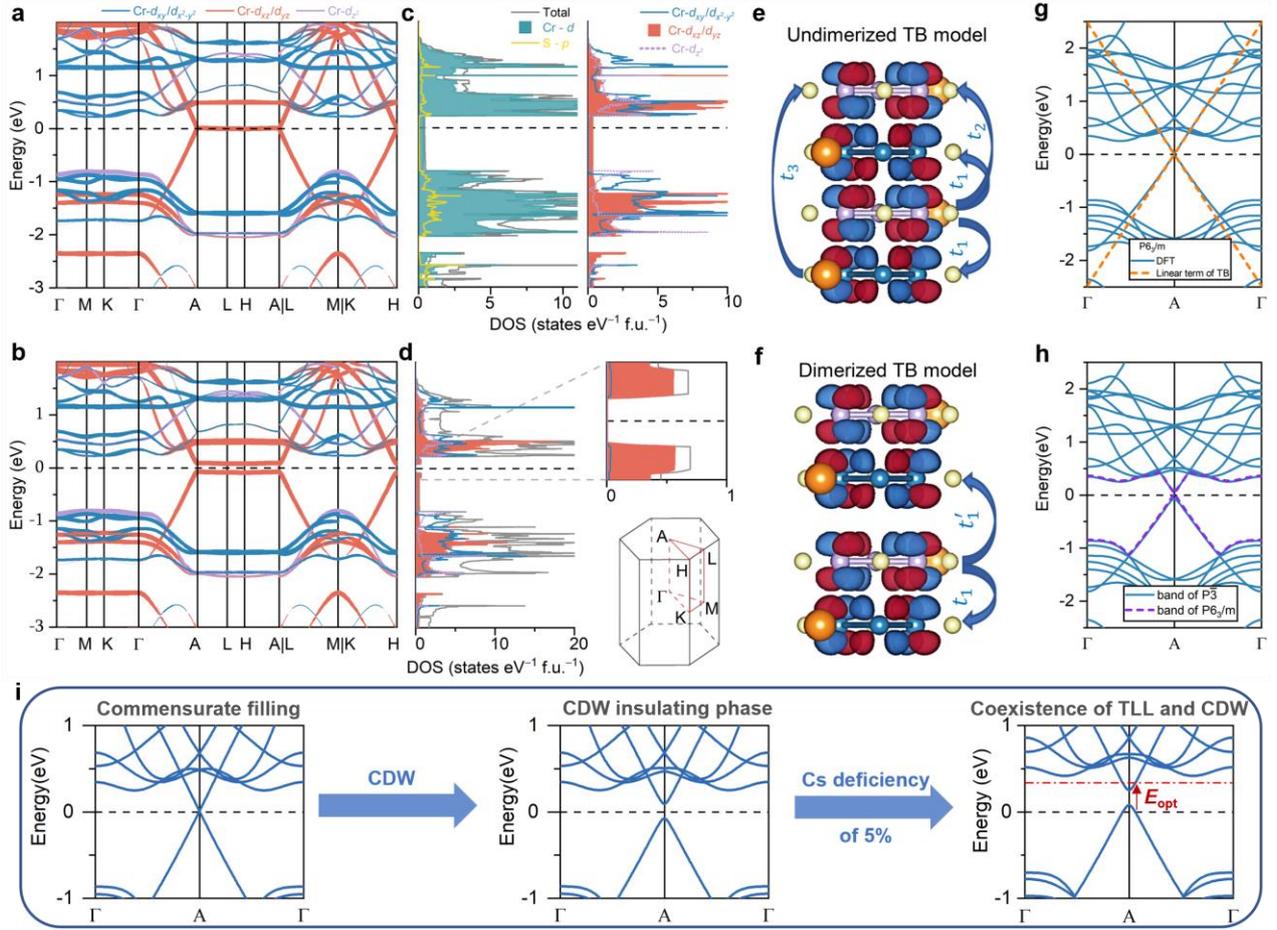

**Figure 4. Electronic-structure perspective on CDW and TLL behaviour. a, b**, Band structures for undimerized and dimerized $CsCr_3S_3$ structures, respectively. The contributions from Cr-3$d$ orbitals are labelled by individual colours. **c,** DOS projected onto S-3$p$ and Cr-3$d$ orbitals for undimerized $CsCr_3S_3$. **d**, The Cr orbital-projected DOS for dimerized structure. The top right panel depicts enlarged views of DOS near the Fermi level. The bottom right panel shows Brillouin zone with high-symmetry points marked. **e, f,** Real-space distributions of $C_3$-symmetric molecular orbitals (MOs) for two sublattices from the linear combination of $d_{yz/xz}$ orbitals. Blue and red colours in the MO stand for opposite signs of phases. **g,** The orange dashed lines represent the dispersion along Γ−A−Γ obtained by retaining only the linear part of the TB model (considering only $t_1$ and $t_3$ hoppings). **h,** Comparison of DFT bands between dimerized and undimerized one near $E_F$. **i**, Evolution of band structures along Γ−A−Γ, interpreting emergent TLL from the electronic bands of a CDW phase. The red arrow on the right panel represents an optical excitation.

## Methods

### Single crystal growth

Employing high-throughput calculations of formation energies in materials prediction within the $KCr_3As_3$ family, we identified a novel chromium chalcogenide. Single crystals (dimensions up to 0.2 × 0.2 × 6 mm$^3$, Extended Data Fig. 1a) of $Cs_{1-\delta}Cr_3S_3$ were grown using CsCl as flux. High purity of cesium (Alfa, 99.8%), presynthesized (solid-state reaction in vacuum between chromium and sulfur at 750 °C for 24 h) chromium sulfide, and cesium chloride (Aladdin, 99.99%, dried at 400 °C for 24 h) were mixed in the molar ratio of 1:3:10. The mixture was loaded into an alumina crucible and sealed inside a Ta tube by arc welding under argon atmosphere. The Ta tube was further sealed in an evacuated silica ampoule and transferred into a muffle furnace. The sealed ampoule was slowly heated to 950 °C, holding for 12 h, and then cooled down to 600 °C at a rate of 2 °C/h. The crystals were harvested by immersing the melts in ethanol for 24 h.

### Structural characterization

Single-crystal X-ray diffraction (SXRD) was performed using a Bruker D8 Venture diffractometer with Mo $K_\alpha$ radiation. A small crystal of $Cs_{1-\delta}Cr_3S_3$ was mounted on the sample holder using the polybutenes oil. Data reduction, including integration and scaling, was performed using the commercial software package APEX4. High-resolution transmission electron microscopy (HRTEM) was conducted using a FEI Tecnai G$^2$ F20 S-Twin transmission electron microscope at room temperature. The chemical composition was checked by energy-dispersive X-ray spectroscopy (EDS) in a field-emission scanning electron microscope operated at an accelerating voltage of 20 kV.

### Physical property measurements

The measurements of electrical transport, heat conductivity and specific heat were carried out on a physical property measurement system (PPMS-9, Quantum Design). The resistance of $Cs_{1-\delta}Cr_3S_3$ is approximately 1 kΩ at room temperature and increases to nearly 1 MΩ upon cooling, far exceeding the load resistance (~10 Ω). To enable reliable resistivity measurements at low temperatures, we employed a two-terminal method with an a.c. excitation of 0.2 V (17.317 Hz) from a lock-in amplifier (SR830), ensuring the measurement within the Ohmic regime. The *I–V* characteristics were also measured in a two-terminal configuration using a current source (Keithley 2400). The thermal conductivity was measured using a steady-state technique. A temperature gradient of ~0.2 K/mm along the chain direction at low temperatures was established using a micro-heater, and precisely monitored by a pair of type-E differential thermocouples, with the heat flux determined from the applied heating power. The specific heat was determined using a thermal relaxation method. Dozens of the crystals (total mass was 1.0 mg) were affixed to the heat capacity puck with N-grease. Data for addenda were collected in advance to ensure accuracy.

### Synchrotron-based ARPES measurements

The ARPES measurements were performed at the BLOCH beamline in Max IV Lab (Sweden) and the BL03U beamline at Shanghai Synchrotron Radiation Facility (China), both equipped with Scienta Omicron DA30 electron analysers with deflection scan capabilities. The typical energy resolution was set to ~20 meV. The photon energy dependent measurement was conducted using photon energies between 60 eV and 100 eV, covering over one Brillouin zone along $k_y$. The measurement temperature ranged from 20 K to 120 K. The typical beamspot is ~15 × 15 μm² for the BLOCH beamline and ~50 × 50 μm² for the BL03U beamline. The single-crystal samples were cleaved *in-situ* under ultrahigh vacuum (< 1×10$^{-10}$ mbar), with the *c* axis (the chain direction) lying in-plane. All ARPES data in the manuscript were collected with vertically polarized photons. The linear dichroism of spectral weight agrees well with orbital projections of each band (Extended Data Fig. 8).

**Density Function Theory calculations**
The first-principles calculations are performed with density functional theory (DFT) method, as implemented in the Vienna Ab initio Simulation Package package (VASP)[35], using the projected augmented wave (PAW) method[36]. The exchange-correlation energy was treated with the Perdew-Burke-Ernzerhof (PBE) generalized gradient approximation (GGA) [37]. An energy cutoff of 600 eV and a Γ-centered 6 × 6 × 16 *k*-point mesh were employed, after careful convergence tests. Full structural optimization was performed starting from the experimental high-temperature crystal structure, including both lattice constants and atomic coordinates. The relaxed structure with lattice parameters of *a* = 9.29 Å and *c* = 4.15 Å, satisfying the experimental values within 98% and 99% respectively. Based on the results of homogeneous Peierls distortion, the Virtual Crystal Approximation (VCA) method was employed to mimic the Cs deficiency. The calculated density of state for dimerized $Cs_{1-\delta}Cr_3S_3$ (δ = 0.05) is 0.69 eV$^{-1}$ f.u.$^{-1}$ at Fermi level. To examine the dynamical stability, we calculated the phonon spectra based on the optimized structure using the Phonopy package in conjunction with the finite displacement method[38,39]. Unlike the PBE functional employed for the electronic structure, the PBEsol exchange-correlation functional was specifically adopted for the phonon calculations to provide more accurate lattice dynamics. We constructed the tight-binding model with the Wannier90 package[40] (detailed in minimal tight-binding model below). Consistent with magnetic measurements (Extended Data Fig. 3), the GGA calculations for representative magnetic orderings (CLK, FM, IAF and IOP)[41] all relax to non-magnetic state with zero magnetic moments. Therefore, the non-magnetic configuration was adopted in all calculations.

**Low-temperature thermal conductivity in a clean TLL state**
The quantized thermal conductance of a single one-dimensional (1D) channel in the absence of impurity backscattering was derived by Kane and Fisher as[32]

$$G_Q = \frac{\pi^2 k_B^2}{3h} T,$$

where $G_Q$ is the thermal conductance (W K$^{-1}$) of an individual ballistic channel. In this context, each 1D conduction channel contributes a universal quantum of thermal conductance proportional to *T*. For a macroscopic quasi-1D solid composed of many parallel chains, the total thermal conductivity can be expressed as

$$\kappa = \frac{N l_{\text{eff}}}{S} G_Q,$$

where *N/S* is the areal density of independent 1D channels, and $l_{\text{eff}}$ is the effective backscattering-free transport length. This relation is derived from the definition of thermal conductance, $G = \kappa S/l$, which is directly analogous to the correspondence between electrical conductance and electrical conductivity.

The effective transport length can be estimated as $l_{\text{eff}} \approx \eta L$ ~4 μm , where *η* ~0.008 is the fitted junction ratio from the TLL scaling in Fig. 2d, and $L$ is the macroscopic sample length (0.5 mm). The crystallographic arrangement of the chains, with an inter-chain spacing of 9.09 Å forming a 120° parallelogram lattice, yields an areal chain density $N/S \approx 1.4 \times 10^{18}$ m$^{-2}$. Substituting these values into the above expressions gives a theoretical thermal conductivity coefficient, $\kappa/T = A_{\text{cal}} = [\pi^2 k_B^2/(3h)](N/S)l_{\text{eff}} \approx 5.3$ W m$^{-1}$ K$^{-2}$, which agrees in magnitude with the experimental value of $A = 2.4$ W m$^{-1}$ K$^{-2}$.

**Minimal tight-binding model**

Given that the energy bands at around $E_F$ are dominantly derived from $3d_{yz/xz}$ orbitals of Cr atoms, as revealed from the DFT calculation, we therefore construct two molecular orbitals through linear combination of the atomic orbitals,

$$|MO_1\rangle = \frac{1}{\sqrt{3}}\left[d_{yz/xz}(Cr_{1a}) + d_{yz/xz}(Cr_{1b}) + d_{yz/xz}(Cr_{1c})\right]$$

$$|MO_2\rangle = \frac{1}{\sqrt{3}}\left[d_{yz/xz}(Cr_{2a}) + d_{yz/xz}(Cr_{2b}) + d_{yz/xz}(Cr_{2c})\right]$$

The two-band Hamiltonian reads,

$$H_t(k) \sim \begin{bmatrix} 2t_2 \cos k_z c & t_1(1 + e^{ik_z c}) + t_3(e^{-ik_z c} + e^{2ik_z c}) \\ t_1(1 + e^{-ik_z c}) + t_3(e^{ik_z c} + e^{-2ik_z c}) & 2t_2 \cos k_z c \end{bmatrix}$$

It can be expanded in terms of the identity ($\sigma_0$) and Pauli matrices ($\sigma_1, \sigma_2, \sigma_3$) as,

$$H_t(k) = \epsilon_0(k)\sigma_0 + \epsilon_1(k)\sigma_1 + \epsilon_2(k)\sigma_2 + \epsilon_3(k)\sigma_3$$

where $\epsilon_i$ (i = 0, 1, 2, 3) are real-valued functions of the crystal momentum $k$. The calculation result indicates that the nearest hopping is dominant (Extended Data Tables 4, 5). Near the A point, $k_z c = \pi + q$, we get,

$$\epsilon_0(q) = -2t_2 \cos q$$
$$\epsilon_1(q) = t_1(1 - \cos q) + t_3(\cos 2q - \cos q)$$
$$\epsilon_2(q) = t_1 \sin q - t_3(\sin 2q + \sin q)$$
$$\epsilon_3(q) = 0$$

$$E_{\pm}(A + q) = -2t_2 \cos q \pm \sqrt{\left(\frac{t_1}{2} - \frac{3t_3}{2}\right)^2 q^4 + (t_1 - 3t_3)^2 q^2}$$

$$\sim (t_1 - 3t_3)q + t_2 q^2 + \frac{1}{8}(t_1 - 3t_3)q^3 + O(q^4)$$

Similarly, the two-band Hamiltonian for dimerized model reads,

$$H_t(k) \sim \begin{bmatrix} 2t_2 \cos k_z c & t_1 + t_1' e^{ik_z c} + t_3 e^{-ik_z c} + t_3' e^{2ik_z c} \\ t_1 + t_1' e^{-ik_z c} + t_3 e^{ik_z c} + t_3' e^{-2ik_z c} & 2t_2 \cos k_z c \end{bmatrix}$$

Near the A point, $k_z c = \pi + q$, we get,

$$E_{\pm}(A + q) = -2t_2 \cos q \pm \sqrt{\left((t_1 - t_1') + (t_3 - t_3') + \left(\frac{t_1'}{2} + \frac{t_3}{2} - 2t_3'\right)q^2\right)^2 + (t_1' - t_3 - 2t_3')^2 q^2}$$

Based on the effective band dispersion derived by including hopping terms up to $t_3$, we analyse the low-energy limit near $q = 0$. In this limit, the Peierls distortion generates a finite term $\Delta$, which is given by the sum of the differences between inequivalent hopping parameters. The corresponding Peierls gap is then, $2\Delta = 2\sum_{n=0}^{1}|t_{2n+1} - t'_{2n+1}|$.

The minimal tight-binding model for the undimerized state, parameterized by low-order hoppings ($t_1$, $t_2$, $t_3$), produces a linear dispersion in good agreement with DFT bands (Fig. 4g). Upon dimerization, a single energy gap opens in the electronic structure; however, the bands away from the gap largely preserve their linear dispersion (Fig. 4h). Owing to hole doping (~5% Cs deficiency), the Fermi level shifts away from the gap, such that the band at the Fermi level remains nearly linear.

**Peierls distortion in a diatomic AB-type chain**
In CsCr$_3$S$_3$ system, there are two sublattices in a unit cell. The Cr$_3$S$_3$-dominated constituent chains can be modelled by a diatomic AB-type chain. Each unit cell contains two sites and accommodates two electrons, giving rise to a fully filled bonding band in terms of molecular orbitals (MOs). The Fermi wavevector is therefore pinned at the boundary of the BZ without introducing an additional doubling of the real-space periodicity. The electronic instability arises from the intra-unit-cell bonding described with the MOs basis.

To rigorously confirm the band filling, we performed a bonding analysis by computing: i) $k$-resolved Crystal Orbital Hamiltonian Population (COHP) calculated by $\langle\psi_a|\hat{H}|\psi_b\rangle$ for all bands and $k$-points along Γ–A–Γ from two-bands tight-binding Hamiltonians, and ii) real-space Bloch functions obtained from the DFT calculation. As shown in Extended Data Figs. 9a–f, the calculated COHP exhibits a continuous evolution of bonding character across momentum: the lowest state of band 1 is described by $\Psi_1$, which is a purely bonding state; whereas the highest state of band 2, with the wave function of $\Psi_4$, is purely antibonding. This clear bonding-antibonding scenario, corroborated by the spatial distribution of the Bloch functions from the DFT calculation, demonstrates that the occupied states correspond to a fully filled bonding manifold.

The essential distinction between the Peierls distortions in diatomic AB chain and canonical monatomic case originates from their initial BZ constructions, as illustrated in Extended Data Figs. 9 e, f. For a monatomic chain, a half-filled band yields $k_F = \pi/(2a)$. The Peierls instability occurs at $q = 2k_F = \pi/a$, which corresponds to the initial BZ boundary. This instability drives a lattice dimerization (a doubling of the unit cell) and a consequent folding of the BZ boundary to Fermi vector. In the folded band picture, a gap opens at the Fermi level only when the distortion introduces unequal hopping integrals ($t_{1L} \neq t_{1R}$). In contrast, the AB chain is already doubled in the undistorted phase due to the two-atom basis, resulting in an inherently folded BZ of size $\pi/(2a)$. The pre-distortion electronic structure naturally splits into bonding and anti-bonding bands, as demonstrated above. The Peierls instability manifests itself as a bond disproportionation within the existing unit cell ($t_{1L} \neq t_{1R}$), which directly opens a band gap at the BZ boundary where the Fermi vector resides, without any further change in translational periodicity.

To conclude, whereas the monatomic chain undergoes a translational symmetry breaking (unit cell doubling), the diatomic AB chain exhibits a Peierls instability rooted in intra-unit-cell bond symmetry breaking. This explains the gap opening at the pre-folded zone boundary for CsCr$_3$S$_3$.

**Extended Data**

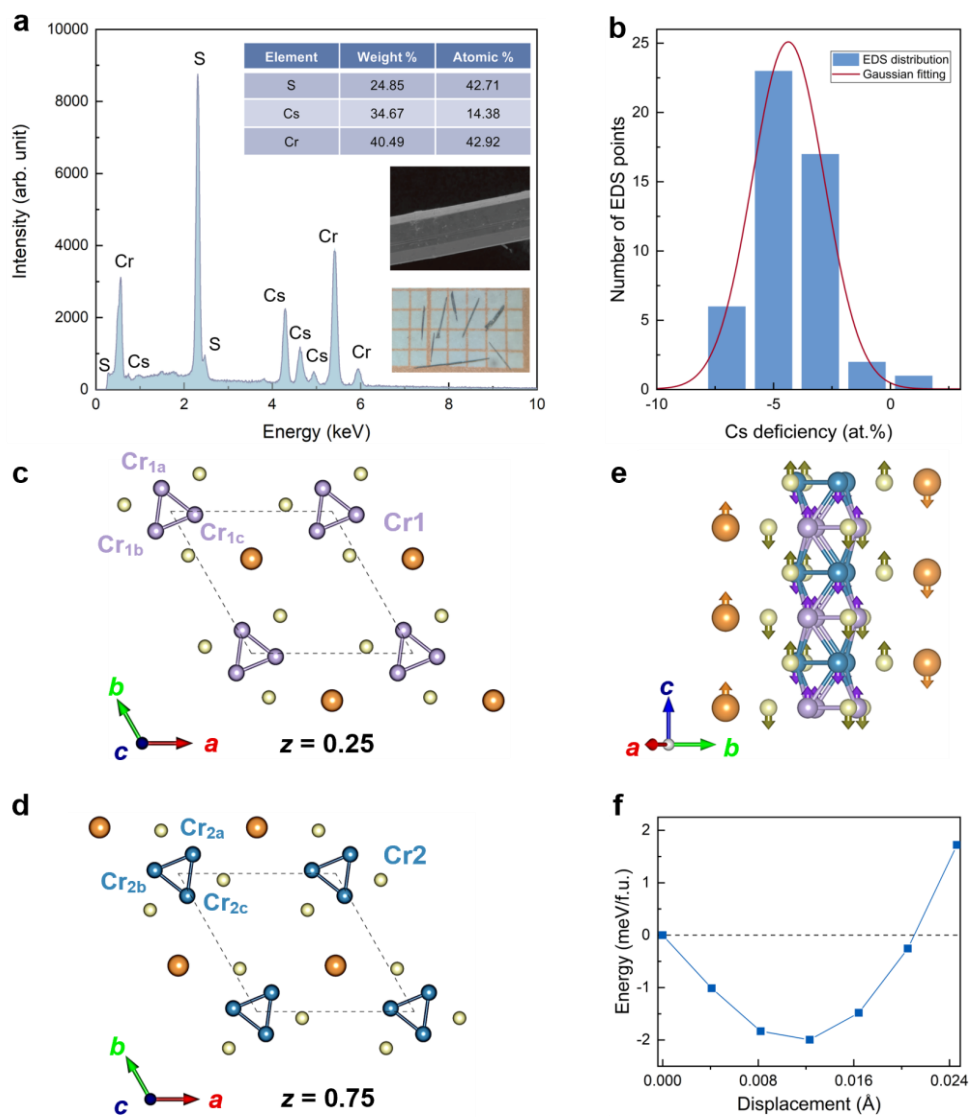

**Extended Data Fig. 1. Compositional and structural characterization of $Cs_{1-\delta}Cr_3S_3$ crystals. a,** A typical EDS spectrum with an SEM image (middle inset) and optical photographs (bottom inset). **b,** Large-scale EDS compositional map showing substantial Cs deficiency with $\delta \sim 0.05$. **c, d,** Top views of the crystal structure at $z = 0.25$ and $z = 0.75$. **e,** Schematic illustration of atomic displacements with the intra-unit-cell dimerization. The arrows show the distortion directions. **f,** Total energy of $CsCr_3S_3$, referenced to that of the undimerized structure, as a function of the displacement of Cr atoms along $c$ axis. An energy minimum is seen at 0.012 Å.

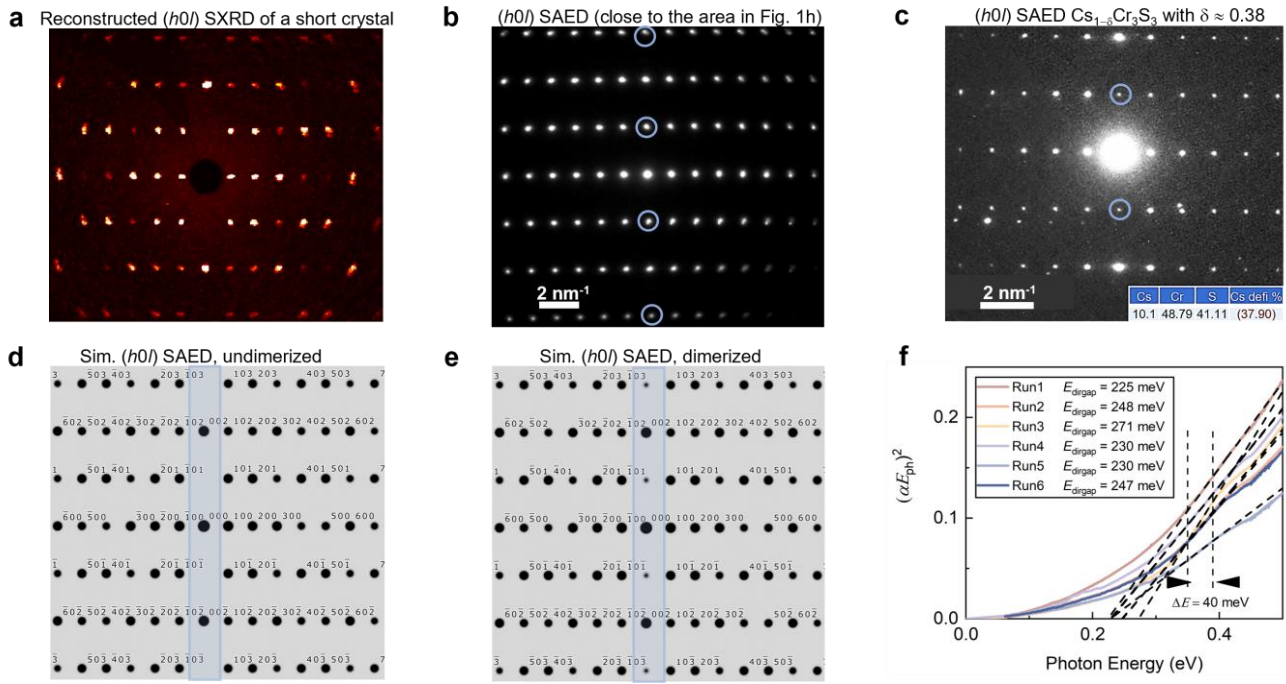

**Extended Data Fig. 2 Experimental and simulated diffraction patterns, and optical absorption spectra. a,** Experimental SXRD diffraction patterns (*h*0*l*) for a short single crystal. **b,** Additional experimental SAED pattern from crystal region near that shown in Fig. 1h. **c,** Experimental SAED pattern of deintercalated $Cs_{1-\delta}Cr_3S_3$ with δ ≈ 0.38 (after soaking in $I_2$/$CH_3CN$ at 50 °C for 24 h). **d, e,** Simulated SAED patterns of the undimerized and dimerized structures in (*h*0*l*). Blue regions indicate the forbidden reflections in the undimerized phase. **f,** Optical absorption spectra of additional samples of $Cs_{1-\delta}Cr_3S_3$ (δ~0.05).

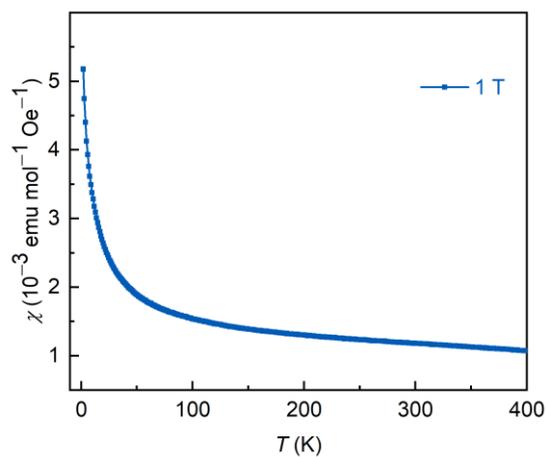

**Extended Data Fig. 3 Temperature dependence of magnetic susceptibility with an applied field of 1T.** There is no indication of any anomaly up to 400 K.

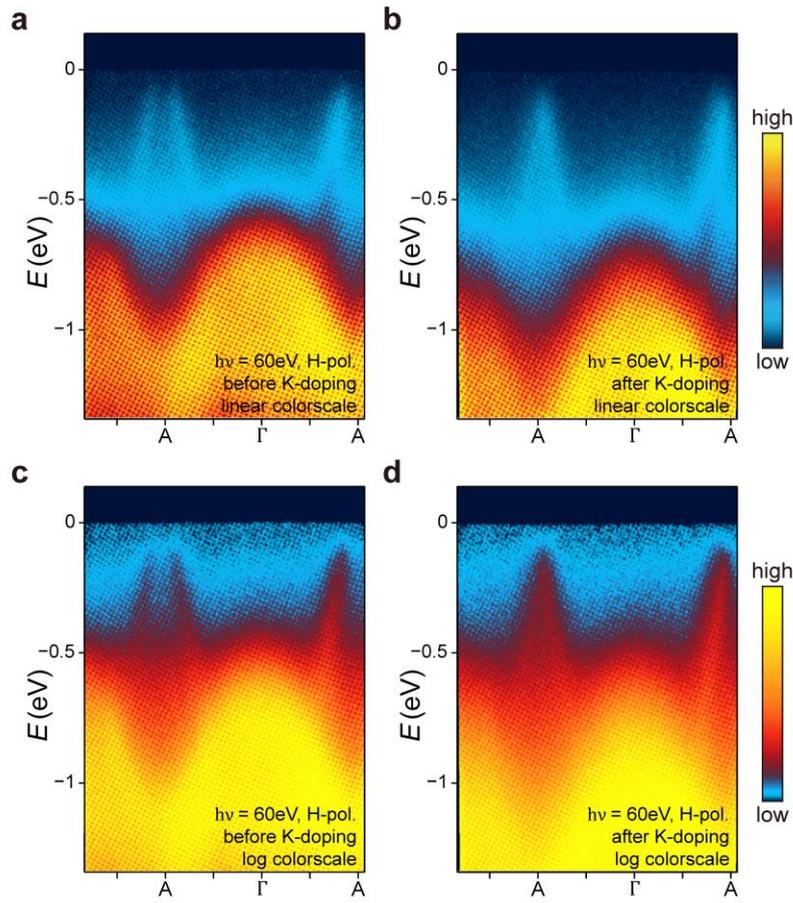

**Extended Data Fig. 4 Effect of *in-situ* potassium doping on the electronic states in $Cs_{1-\delta}Cr_3S_3$. a, b,** ARPES spectrum measured before (a) and after (b) depositing potassium. The sample was cleaved *in-situ* under ultra-high vacuum before potassium doping. **c, d,** The same spectra of **a** and **b** with log colorscale. It is clear that the valence bands shift down after the electron doping, resulting in a band gap near $E_F$. This experiment supports the existence of the gap opening due to dimerization, consistent with the DFT calculation shown in Fig. 4b. Note that the grid patterns here are caused by the fixed scan mode used in these scans (all other scans in this paper were obtained from swept mode).

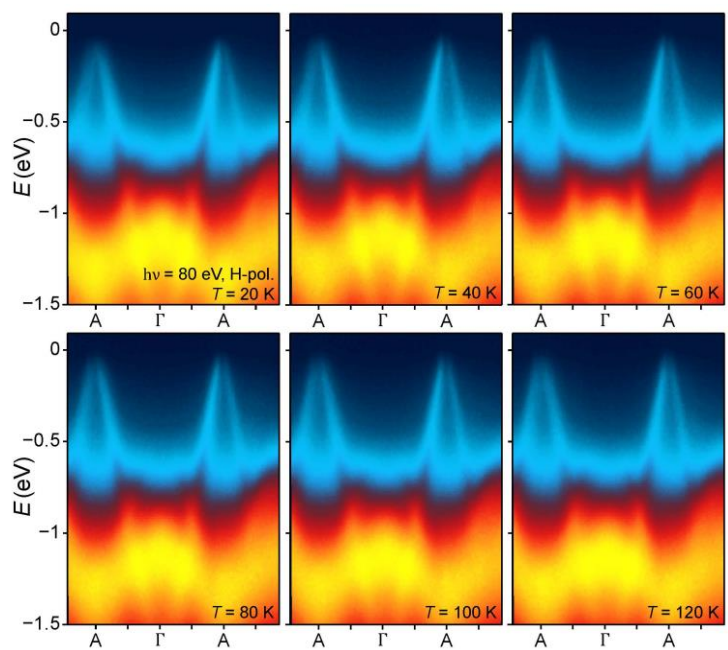

**Extended Data Fig. 5 Temperature-dependent ARPES measurements conducted at temperatures ranging from 20 K to 120 K.**

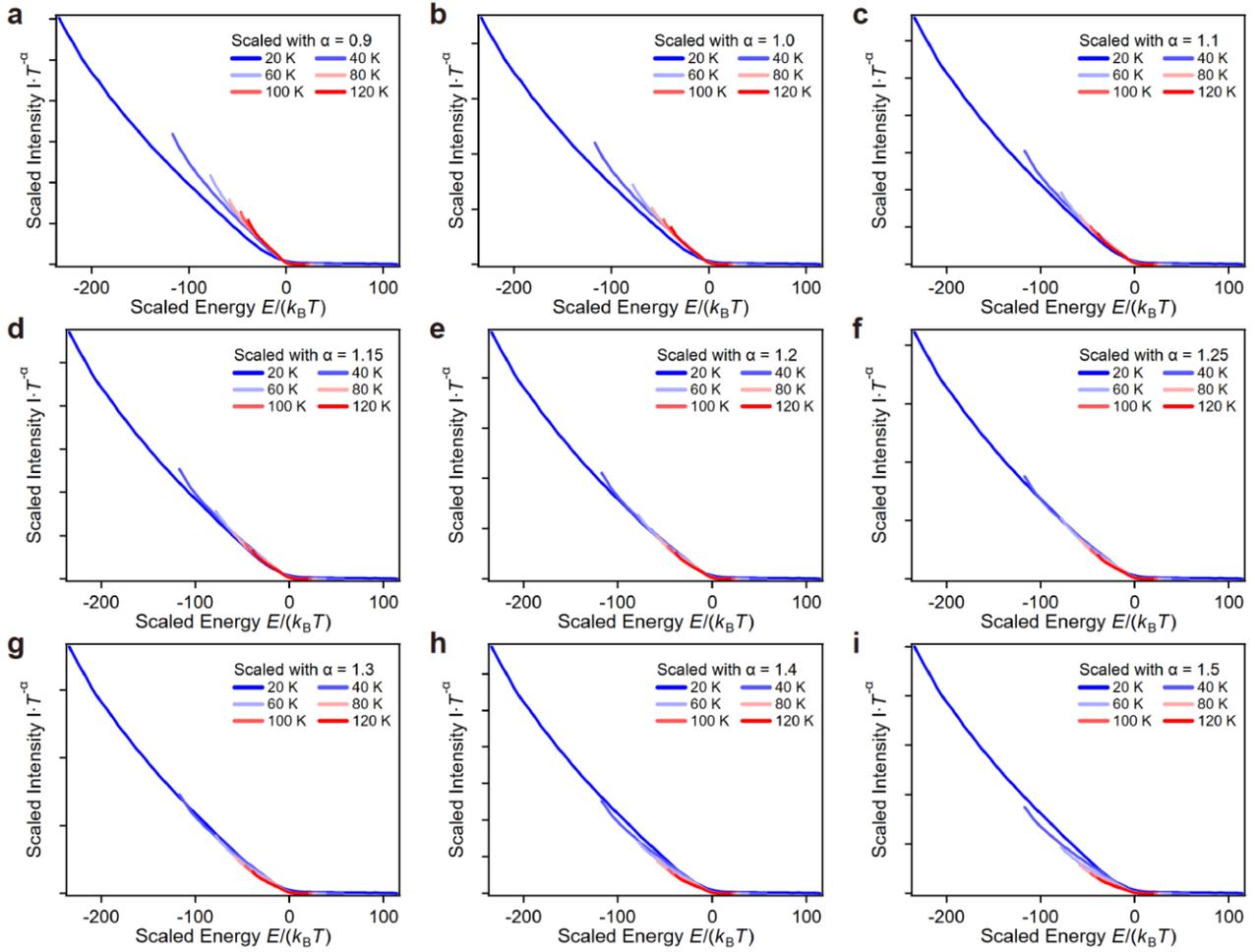

**Extended Data Fig. 6 Scaling of temperature-dependent EDCs [$I/T^{\alpha}$ against $E/(k_{B}T)$] with various values of $\alpha$.** The range of $E$ is −0.4~0.2 eV. The best scaling is achieved at $\alpha = 1.2$.

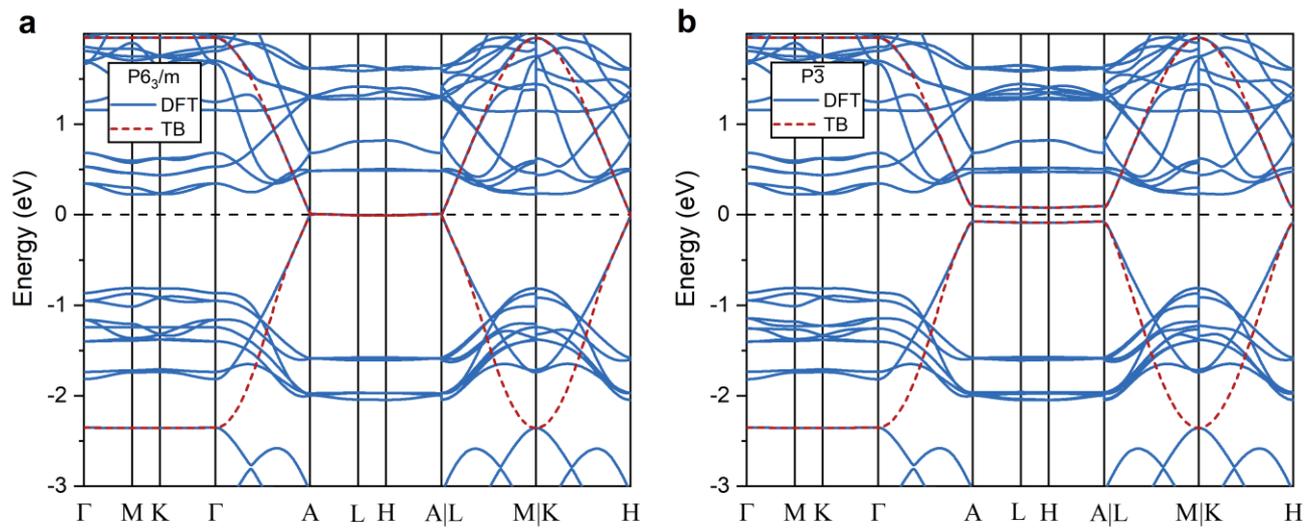

**Extended Data Fig. 7 Comparison of the results of tight-binding model and DFT for the undimerized (a) and dimerized (b) CsCr$_3$S$_3$.**

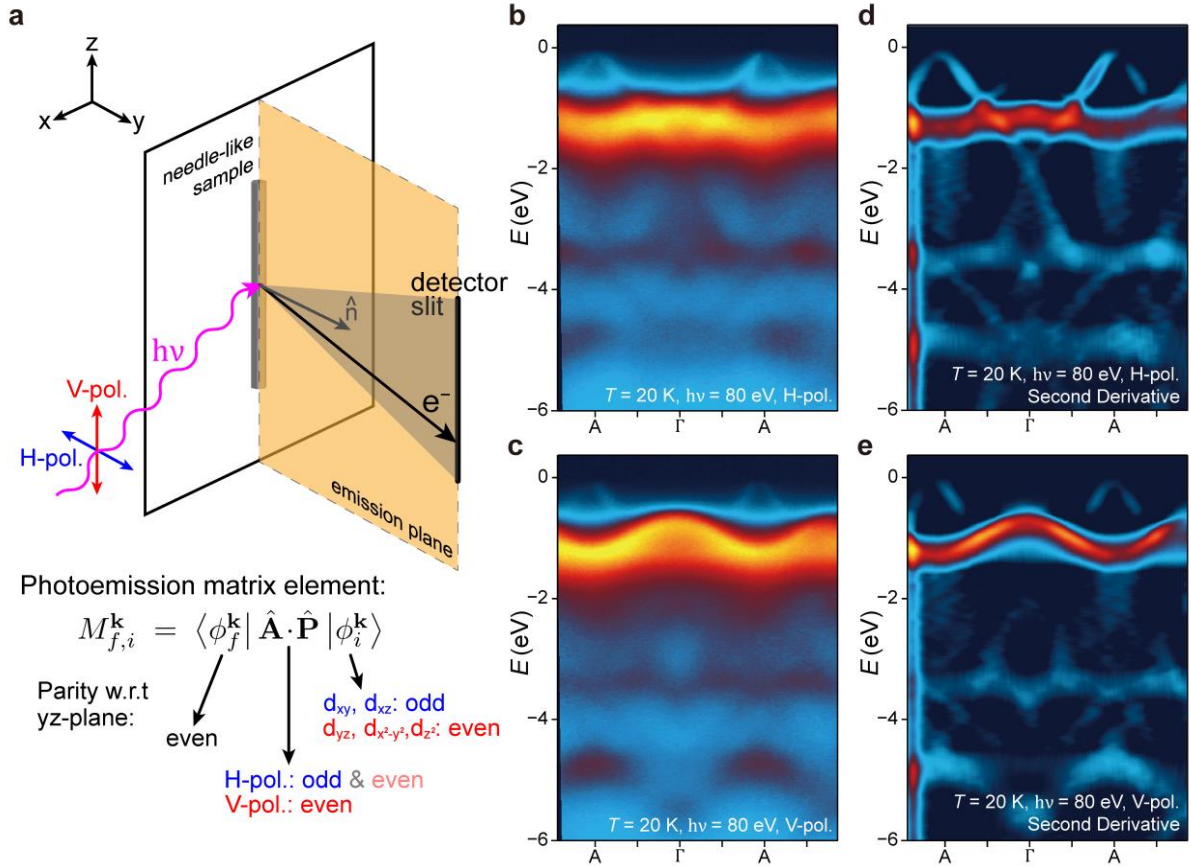

**Extended Data Fig. 8 Polarization-dependent ARPES measurements. a,** ARPES measurement geometry. H/V-pol. refers to the horizontally/vertically polarized photons. **b, c,** ARPES raw data measured with horizontally/vertically polarized (H/V-pol.) photons, respectively. **d, e,** Second derivatives of **b** and **c**. Note that the H-pol. spectrum highlights the linear bands crossing $E_\mathrm{F}$, while V-pol. photons enhance the lower hole band centred at Γ point. Since the linear band originates from Cr-$d_{yz/xz}$ electrons with opposite parities (see Fig. 4b), finite spectral weight can be detected with both H- and V-polarized light. In contrast, the hole pocket at −1 eV derives from the Cr-$d_{z^2}$ orbital and is therefore only visible in the V-polarized measurements.

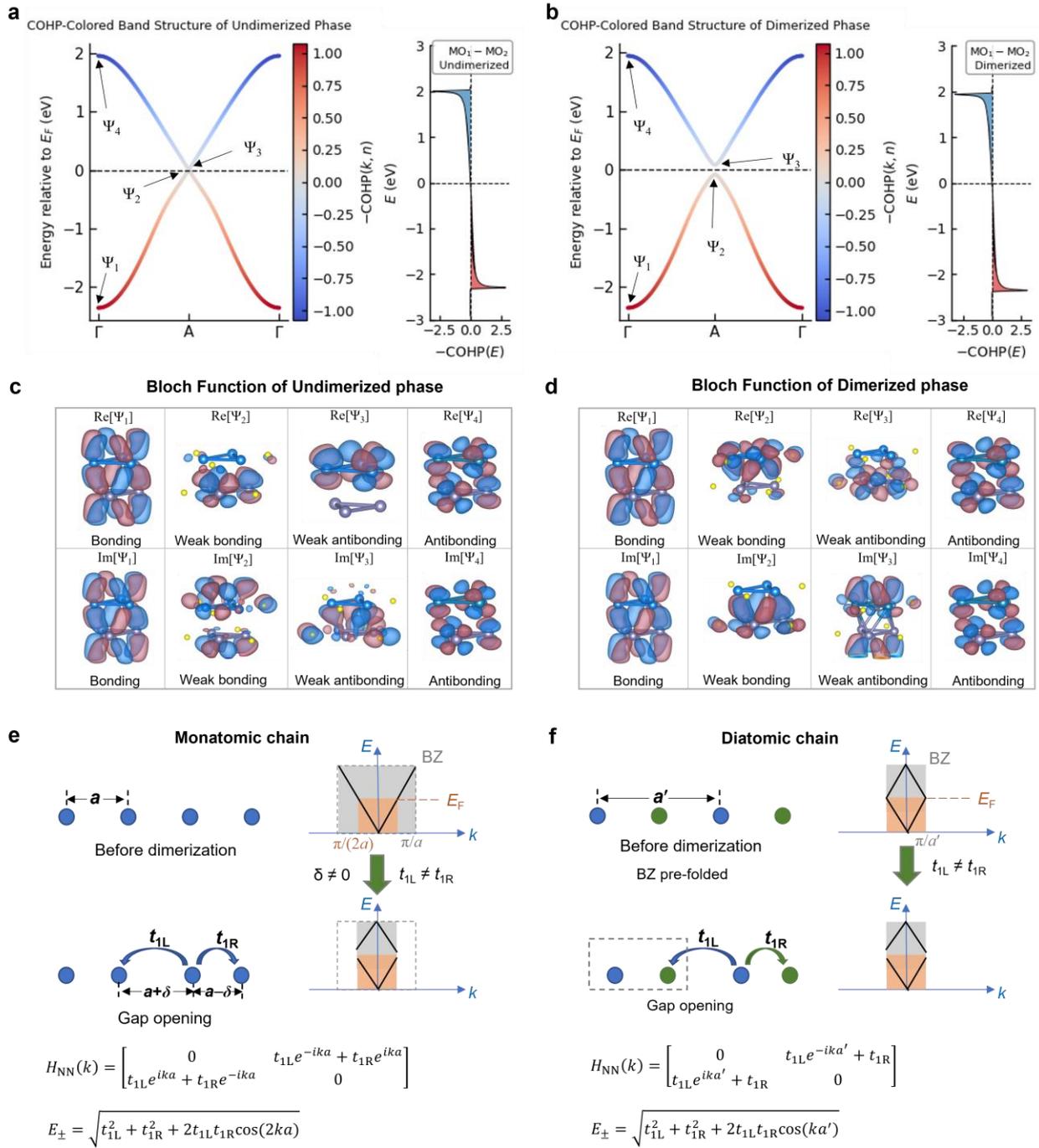

**Extended Data Fig. 9 Bonding-antibonding character in CsCr₃S₃ and origin of the Peierls instability. a,** $k$- and energy-resolved COHP($k$, $n$) ($n$ is the band index) of two tight-binding MOs-derived bands for undimerized CsCr₃S₃, showing bonding/antibonding crossover at $E_F$. **b,** $k$- and energy-resolved COHP for dimerized CsCr₃S₃. **c, d,** Real-space amplitudes of representative Bloch states $\Psi_1$, $\Psi_2$, $\Psi_3$, and $\Psi_4$, corresponding to the $k$ points marked in **a** and **b**, respectively. **e, f,** Schematic comparison of Peierls distortions in monatomic and diatomic chains. The tight-binding Hamiltonians and their corresponding eigenvalues are given, and the evolutions of the band structures across the Peierls transition are displayed.

**Extended Data Table 1.** TLL parameter $K$ extracted from different experimental probes in this work, benchmarked against canonical TLL/Peierls systems. Some of the $K$ values are derived from the reported exponents ($\alpha$) using the relation, $\alpha = \frac{K+K^{-1}-2}{4}$. In $A_{0.3}$MoO$_3$, Perierls instability (or CDW order) emerges at low temperature, where the TLL behaviour disappears.

| Systems | Reported properties | Method | $K$ | Refs. |
|---|---|---|---|---|
| Cs$_{1-\delta}$Cr$_3$S$_3$ | TLL and Peierls | $R$–$T$ | 0.11 | (Our work) |
| | | $I$–$V$ | 0.12 | |
| | | ARPES | 0.15 | |
| K$_2$Cr$_3$As$_3$ | TLL to superconductivity | ARPES | 0.17 | Ref. 14 |
| Tl$_2$Mo$_6$Se$_6$ | TLL to superconductivity | ARPES | 0.3 | Ref. 42 |
| Ta$_2$Pd$_3$Te$_5$ | TLL edge state | d$I$/d$V$ | 0.14–0.53 | Ref. 43 |
| Li$_{0.9}$Mo$_6$O$_{17}$ | TLL | $R$–$T$ | 0.25 | Ref. 44 |
| | | STM | 0.24 | Ref. 10 |
| $A_{0.3}$MoO$_3$ | TLL to Peierls | ARPES | 0.24 | Ref. 45 |
| Na$_{2-\sigma}$Mo$_6$Se$_6$ | TLL | $R$–$T$ | 1.5 | Ref. 46 |
| SrNbO$_{3.41}$ | TLL | $R$–$T$ | 0.2 | Ref. 47 |
| MoSe$_2$ | TLL | STM | 0.54 | Ref. 48 |
| tWTe$_2$ 2D moiré lattice | TLL | d$I$/d$V$ | 0.17 | Ref. 21 |
| Carbon nanotube | TLL | $R$–$T$ & $I$–$V$ & d$I$/d$V$ | 0.3 | Ref. 19 |
| CoSb$_{1-x}$ nanoribbons | TLL | ARPES | 0.21 | Ref. 49 |
| MoSe nanowires | TLL | $R$–$T$ & $I$–$V$ | 0.04 (small diameter)–0.23 | Ref. 17 |
| K$_2$Pt(CN)$_4$Br$_{0.3}$·2H$_2$O | Peierls | – | – | Ref. 50 |

**Extended Data Table 2. Crystallographic and structure refinement details for $Cs_{1-\delta}Cr_3S_3$ at 170 K from the single-crystal XRD in space group $P6_3/m$ (No. 176).**

| | |
|---|---|
| Formula | $Cs_{0.956}Cr_3S_3$ |
| Formula weight | 379.24 |
| Temperature | 170 K |
| Wavelength | 0.71073 Å |
| Crystal system | Hexagonal |
| Space group | $P6_3/m$ |
| Unit cell dimensions | $a$ = 9.0915(3) Å, $\alpha$ = 90° |
| | $b$ = 9.0915(3) Å, $\beta$ = 90° |
| | $c$ = 4.1525(2) Å, $\gamma$ = 120° |
| Volume | 297.24(2) Å$^3$ |
| Z | 2 |
| Density (calculated) | 4.237 g/cm$^3$ |
| Absorption coefficient | 12.058 mm$^{-1}$ |
| $F(000)$ | 345 |
| Crystal size | 0.12 × 0.31 × 2.00 mm$^3$ |
| $\theta$ range for data collection | 2.587 to 30.491° |
| Index ranges | $-12 \leq h \leq 12, -12 \leq k \leq 12, -5 \leq l \leq 5$ |
| Reflections collected | 14139 |
| Independent reflections | 341 [$R_{int}$ = 0.0312] |
| Completeness to $\theta$ = 25.242° | 100% |
| Refinement method | Full-matrix least-squares on $F^2$ |
| Data / restraints / parameters | 341 / 0 / 16 |
| Goodness-of-fit | 0.767 |
| Final $R$ indices [$I > 2\sigma(I)$] | $R_{obs}$ = 0.0173, $wR_{obs}$ = 0.0765 |
| $R$ indices [all data] | $R_{all}$ = 0.0173, $wR_{all}$ = 0.0765 |
| Extinction coefficient | 0.031(5) |
| Largest diff. peak and hole | 1.330 and $-0.922$ eÅ$^{-3}$ |

\* $R = \Sigma||F_o|-|F_c|| / \Sigma|F_o|$, $wR = \{\Sigma[w(|F_o|^2 - |F_c|^2)^2] / \Sigma[w(|F_o|^4)]\}^{1/2}$ and $w = 1/[\sigma^2(F_o^2) + (0.1000P)^2]$ where $P = (F_o^2 + 2F_c^2)/3$

| Label | $x$ | $y$ | $z$ | Occupancy | $U_{eq}$* | $U_{11}$ | $U_{22}$ | $U_{33}$ | $U_{12}$ | $U_{13}$ | $U_{23}$ |
|---|---|---|---|---|---|---|---|---|---|---|---|
| Cs | 1/3 | 2/3 | 1/4 | 0.956 | 11(1) | 10(1) | 10(1) | 14(1) | 5(1) | 0 | 0 |
| Cr | 0.1480(1) | 0.1712(1) | 1/4 | 1 | 6(1) | 7(1) | 6(1) | 6(1) | 4(1) | 0 | 0 |
| S | 0.3180(1) | 0.0445(1) | 1/4 | 1 | 8(1) | 9(1) | 9(1) | 8(1) | 5(1) | 0 | 0 |

\* $U_{eq}$ is defined as one third of the trace of the orthogonalized $U_{ij}$ tensor. The unit of the equivalent isotropic and anisotropic displacement parameters is 0.001 Å$^2$.

**Extended Data Table 3. Crystallographic and structure refinement details for $Cs_{1-\delta}Cr_3S_3$ at 170 K from the single-crystal XRD in space group $P\bar{3}$ (No. 147).**

| | |
|---|---|
| Empirical formula | $Cs_{0.956}Cr_3S_3$ |
| Formula weight | 379.24 |
| Temperature | 170 K |
| Wavelength | 0.71073 Å |
| Crystal system | Hexagonal |
| Space group | $P\bar{3}$ |
| Unit cell dimensions | $a = 9.0915(3)$ Å, $\alpha = 90°$ |
| | $b = 9.0915(3)$ Å, $\beta = 90°$ |
| | $c = 4.1525(2)$ Å, $\gamma = 120°$ |
| Volume | 297.24(2) Å$^3$ |
| Z | 2 |
| Density (calculated) | 4.237 g/cm$^3$ |
| Absorption coefficient | 12.058 mm$^{-1}$ |
| $F(000)$ | 345 |
| Crystal size | $0.12 \times 0.31 \times 2.00$ mm$^3$ |
| $\theta$ range for data collection | 2.587 to 30.491° |
| Index ranges | $-12 \leq h \leq 12, -12 \leq k \leq 12, -5 \leq l \leq 5$ |
| Reflections collected | 14169 |
| Independent reflections | 608 [$R_{int} = 0.0311$] |
| Completeness to $\theta = 25.242°$ | 100% |
| Refinement method | Full-matrix least-squares on $F^2$ |
| Data / restraints / parameters | 608 / 0 / 23 |
| Goodness-of-fit | 0.836 |
| Final R indices [$I > 2\sigma(I)$] | $R_{obs} = 0.0167$, $wR_{obs} = 0.0847$ |
| R indices [all data] | $R_{all} = 0.0168$, $wR_{all} = 0.0848$ |
| Extinction coefficient | 0.032(4) |
| Largest diff. peak and hole | 1.443 and $-0.998$ eÅ$^{-3}$ |

* $R = \Sigma||F_o|-|F_c|| / \Sigma|F_o|$, $wR = \{\Sigma[w(|F_o|^2 - |F_c|^2)^2] / \Sigma[w(|F_o|^4)]\}^{1/2}$ and $w = 1/[\sigma^2(F_o^2) + (0.1000P)^2]$ where $P = (F_o^2+2F_c^2)/3$

| Label | $x$ | $y$ | $z$ | Occupancy | $U_{eq}$* | $U_{11}$ | $U_{22}$ | $U_{33}$ | $U_{12}$ | $U_{13}$ | $U_{23}$ |
|---|---|---|---|---|---|---|---|---|---|---|---|
| Cs | 2/3 | 1/3 | 0.2500(1) | 0.956 | 11(1) | 10(1) | 10(1) | 14(1) | 5(1) | 0 | 0 |
| Cr | 0.1480(1) | 0.1713(1) | 0.2502(1) | 1 | 6(1) | 7(1) | 7(1) | 6(1) | 4(1) | 0(1) | 0(1) |
| S | 0.3180(1) | 0.0445(1) | 0.2505(2) | 1 | 8(1) | 9(1) | 9(1) | 8(1) | 5(1) | 0(1) | 0(1) |

* $U_{eq}$ is defined as one third of the trace of the orthogonalized $U_{ij}$ tensor. The unit of the equivalent isotropic and anisotropic displacement parameters is 0.001 Å$^2$.

**Extended Data Table 4. Hopping matrix elements of the tight-binding Hamiltonian for CsCr$_3$S$_3$ in the real-space representation of undimerized structure.** Indices "A" and "B" refer to the two inequivalent sites within each unit cell, and $\delta R$ denotes the relative displacement between unit cells. $E_0$ (5.3691 eV) is the on-site energy, and $t_1$, $t_2$, $t_3$, $t_4$ are hopping amplitudes (eV) corresponding to increasing inter-site distance (see Fig. 4e).

| sites | AA | AB | BA | BB |
|---|---|---|---|---|
| $\delta R = -2$ | $t_4$ | $t_3$ | $t_5$ | $t_4$ |
| $\delta R = -1$ | $t_2$ | $t_1$ | $t_3$ | $t_2$ |
| $\delta R = 0$ | $E_0$ | $t_1$ | $t_1$ | $E_0$ |
| $\delta R = 1$ | $t_2$ | $t_3$ | $t_1$ | $t_2$ |
| $\delta R = 2$ | $t_4$ | $t_5$ | $t_3$ | $t_4$ |
| Hopping term | $t_1$ | $t_2$ | $t_3$ | $t_4$ |
| value | −1.0085 | −0.0576 | −0.0761 | −0.0020 |
| | Linear | quadratic | cubic | quintic |
| form | $t_1-3t_3$ | $t_2$ | $\dfrac{t_1-3t_3}{8}$ | |
| value | 0.78 | 0.05 | 0.098 | 0.006 |

**Extended Data Table 5. Hopping matrix elements of the tight-binding Hamiltonian for CsCr$_3$S$_3$ in the real-space representation of dimerized structure.** Indices "A" and "B" refer to the two inequivalent sites within each unit cell, and $\delta R$ denotes the relative displacement between unit cells. $E_0$ (5.2849 eV) is the on-site energy, and $t_1$, $t_1'$, $t_2$, $t_3$, $t_3'$, $t_4$ are hopping amplitudes (eV) corresponding to increasing inter-site distance, where both $t_1$ and $t_1'$ represent nearest-neighbour hoppings (see Fig. 4f).

| sites | AA | AB | BA | BB |
|---|---|---|---|---|
| $\delta R = -2$ | $t_4$ | $t_3'$ | $t_5$ | $t_4$ |
| $\delta R = -1$ | $t_2$ | $t_1'$ | $t_3$ | $t_2$ |
| $\delta R = 0$ | $E_0$ | $t_1$ | $t_1$ | $E_0$ |
| $\delta R = 1$ | $t_2$ | $t_3$ | $t_1'$ | $t_2$ |
| $\delta R = 2$ | $t_4$ | $t_5$ | $t_3'$ | $t_4$ |

| Hopping term | $t_1$ | $t_1'$ | $t_2$ | $t_3$ | $t_3'$ | $t_4$ |
|---|---|---|---|---|---|---|
| value | −1.0674 | −0.9850 | −0.0532 | −0.0568 | −0.0582 | 0.0104 |

**Acknowledgments**
We thank Wu-Zhang Yang and Zhi Ren for the assistance with the specific heat measurements. We also thank Dr. Balasubramanian Thiagarajan, Dr. Mats Leandersson, Dr. Zhengtai Liu and Dr. Mao Ye for experimental support during the synchrotron ARPES measurements, which were carried out at BLOCH beamline at MAX IV laboratory and beamline 03U at Shanghai synchrotron radiation facility. This work was supported by the National Key Research, Development Program of China (2022YFA1403202, 2023YFA1406101 and 2023YFA1406303), and the National Natural Science Foundation of China (Grant No. 12404159, 12525408 and 12274283).

**Author contributions**
G.-H.C. coordinated the work and interpreted the results in discussion with J.L., G.-W.Y., S.-Q.W., H.J., C.C., Ya.L., and H.-Q.L. The title material was discovered by J.L. and B.-Z.L, with the help from high-throughput material predictions developed by H.J. The crystal structures were analysed by J.L., Yi.L., J.-K.B., J.-Y.Liu, and H.X.L. The physical property measurements were done by J.L., Yi.L, X.-x.Y., and H.X. The ARPES measurements were carried out by G.-W.Y., supervised by Ya.L. The theoretical calculations were done by J.L. and S.-Q.W., supervised by C.C., G.-H.C. and H.-Q.L. The paper was written by J.L., G.-W.Y., Ya.L., and G.-H.C. All authors commented on the paper.

**Competing interests**
The authors declare no competing interests.

**Correspondence and requests for materials** should be addressed to G.H.C.

**Data availability**
All data are available in the main text or the supplementary materials.